# High-Performance Atomically-Thin Room-Temperature NO$_2$ Sensor

Amin Azizi[1,2], Mehmet Dogan[1,3], Hu Long[1,2,3], Jeffrey D. Cain[1,2,3], Kyunghoon Lee[1,2,3], Rahmatollah Eskandari[1], Alessandro Varieschi[1], Emily C. Glazer[1], Marvin L. Cohen[1,3], Alex Zettl*[,1,2,3]

1. Department of Physics, University of California at Berkeley, Berkeley, CA, 94720, USA
2. Kavli Energy NanoScience Institute at the University of California, Berkeley, Berkeley, CA, 94720, USA
3. Materials Sciences Division, Lawrence Berkeley National Laboratory, Berkeley, CA, 94720, USA

* Corresponding Author. E-mail: azettl@berkeley.edu , Phone: +1 (510) 642-4939   Fax: +1 (510) 642-2685

Abstract:

The development of room-temperature sensing devices for detecting small concentrations of molecular species is imperative for a wide range of low-power sensor applications. We demonstrate a room-temperature, highly sensitive, selective, stable, and reversible chemical sensor based on a monolayer of the transition metal dichalcogenide Re$_{0.5}$Nb$_{0.5}$S$_2$. The sensing device exhibits thickness dependent carrier type, and upon exposure to NO$_2$ molecules, its electrical resistance considerably increases or decreases depending on the layer number. The sensor is selective to NO$_2$ with only minimal response to other gases such as NH$_3$, CH$_2$O, and CO$_2$. In the presence of humidity, not only are the sensing properties not deteriorated, but also the monolayer sensor shows complete reversibility with fast recovery at room temperature. We present a theoretical analysis of the sensing platform and identify the atomically-sensitive transduction mechanism.

Chemical sensing devices are crucial to monitoring environmental pollution, health conditions, and industrial processes. In particular, the detection of nitrogen dioxide ($NO_2$), a toxic gas emitted from power plants, vehicles, and industrial sources, is of significant importance. $NO_2$ can have major environmental effects such as contributing to the creation of acid rain, the formation of ground-level ozone, and catalyzing small particles that trigger a range of health problems. For example, exposure to a high concentration of $NO_2$ can worsen respiratory diseases[1]. In addition to monitoring applications, ultrasensitive sensors can be used for medical diagnoses, such as identifying asthma[2]. Thus, there is a need for more sensitive $NO_2$ sensors (at the parts per billion (ppb) level) that are both selective and reversible. Common sensor materials are metal oxides[3–5], conducting polymers[6,7], and low-dimensional materials[8–11]. Semiconducting metal oxides have been extensively used for detecting $NO_2$. However, they usually must operate at high temperatures to achieve suitable performance[4,5,12]. This hinders their application due to the increased power consumption.

Two-dimensional (2D) materials[13–20] offer extremely high surface-to-volume ratio, layer-dependent electronic structure, and tunable active sites for redox reactions, making them attractive candidates for gas sensing applications[11,21–29]. While they have been widely used for gas sensing applications with an excellent sensitivity for toxic gases such as $NO_2$, the reported ultra-thin $NO_2$ sensors usually suffer from slow recovery kinetics at room temperature, inferior selectivity, and/or poor stability. For example, graphene[21] and $MoS_2$[22,23] have shown excellent sensitivities toward $NO_2$, but they are not selective and suffer from slow recovery kinetics. Black phosphorus has also exhibited an excellent sensitivity for $NO_2$ down to the ppb level[24]. However, its poor stability and selectivity limit its practical application[24]. $SnS_2$ has been shown to be highly sensitive to $NO_2$ with superior selectivity[25]. It also offers excellent reversibility at 120°C, but unfortunately does not exhibit acceptable response and recovery times at temperatures below 80°C[25]. MoTe$_2$ has been demonstrated to be an ultrasensitive NO$_2$ sensor with enhanced sensitivity and recovery rate under ultraviolet illumination, but it suffers from long recovery time in ambient conditions without an external stimulus[29]. Recently, NbS$_2$ has shown excellent selectivity toward NO$_2$[30]. However, its slow response and recovery kinetics at room temperature[30] restrain its application. Furthermore, practical sensors should operate not only at room temperature[31], but also under conditions of high relative humidity. Humidity greatly deteriorates the sensing properties of metal oxide sensors[32,33] and, to a lesser degree, graphene-based sensors[34]. Therefore, the realization of an ultrasensitive, selective, and reversible NO$_2$ sensor that can operate in ambient conditions (*i.e.* at room temperature and under substantial relative humidity) remains elusive.

Here, we explore the NO$_2$ sensing behavior of atomically-thin specimens of Re$_{0.5}$Nb$_{0.5}$S$_2$. We find that the sensing properties depend sensitively on layer number, with monolayers giving the best overall performance. The monolayer is capable of detecting ultralow concentrations of NO$_2$ while showing excellent selectivity. Critically, in the presence of humidity, the monolayer sensor is highly reversible at room temperature with fast response and recovery times. In addition to being highly sensitive, selective, and reversible, the sensor is also stable with no sign of degradation after several cycles of measurements in air with different levels of humidity. We explore theoretically the sensor operation for different gas species using density functional theory (DFT) and identify the underlying sensing mechanism.

Bulk crystals of Re$_{0.5}$Nb$_{0.5}$S$_2$ are synthesized using chemical vapor transport. The samples are mechanically cleaved to atomic thinness. Figure 1a shows an annular dark-field scanning transmission electron microscope (ADF-STEM) image of the monolayer Re$_{0.5}$Nb$_{0.5}$S$_2$. Owing to the Z (atomic number) contrast of the atomic-resolution image, Re atoms (heavier) show a brighter contrast while Nb atoms (lighter) appear dimmer. In addition, the ADF-STEM image in Fig. 1b reveals an AA´ stacking sequence for bilayer and trilayer regions. Chemical composition of the flake is confirmed from the atomically-resolved ADF-STEM image and complementary energy-dispersive X-ray spectroscopy (EDS) measurements (Fig. 1c and Fig. S1), with near-equal concentration of Re and Nb, as expected for $Re_{0.5}Nb_{0.5}S_2$.

$Re_{0.5}Nb_{0.5}S_2$ offers a broad range of bandgap energies, ranging from ~1.03eV to ~0.36eV as the thickness increases from monolayer to bulk[35]. This can result in striking layer-dependent electrical properties. For our sensor platform, different layer-number field-effect transistor (FET) devices are fabricated with a two-terminal back-gate configuration (Fig. 1d and Fig. S2). Prior to sensing measurements, we measure the transport behavior of the devices in vacuum. Fig. 1e demonstrates the change in the drain current ($I_{ds}$) as a function of source-drain bias ($V_{ds}$) for the monolayer (1L), six-layer (6L), and thirty-layer (30L) devices. The linear characteristic of the $I_{ds}-V_{ds}$ curves implies Ohmic contacts. Sweeping the gate voltage, a distinct layer-dependent transport behavior is seen (Fig. 1f). Interestingly, the carrier type of the devices changes with the layer number, *e.g.* n-type for the monolayer device and p-type for the 6L device. Additionally, as the flakes become thicker, the gate control becomes weaker. For instance, the on/off ratio decreases from ∼$1.4*10^4$ for 1L to ~4 for 6L. The 30L device displays a negligible gate control. This can be attributed to the large modification of the $Re_{0.5}Nb_{0.5}S_2$ electronic structure with thickness[35].

For the sensing measurements, we apply a constant $V_{ds}$ of 1V to the two-terminal FET devices and monitor the changes in their electrical resistances upon exposure to different concentrations of gas species. The $NO_2$ molecule, with an unpaired electron, is a strong oxidizer that withdraws electrons from the conduction band of the sensing material, in contrast to electron donor gases (*e.g.* $NH_3$, $CH_2O$) which donate electrons to the sensing material. Consistent with the electrical transport measurements, we observe that the electrical resistance of the 1L sensor increases, while that of the 6L sensor decreases, upon the exposure to $NO_2$. Figure 2a-b show the responses of the 1L, 6L, and 30L devices to different $NO_2$ concentrations ranging from 50 ppb to 15 ppm in dry air. We observe a thickness-dependent response for the sensors. While the 1L and 6L devices show excellent responses to $NO_2$ even at ultra-low concentrations, the 30L sensor is not very sensitive to $NO_2$. This highlights the importance of surface-to-volume ratio of low-dimensional materials for gas sensing applications. Additionally, the drastic change in the electronic structure of $Re_{0.5}Nb_{0.5}S_2$ with the layer number can be another cause of this behavior[35]. We note that metallic single-walled carbon nanotubes (SWNTs) typically show small resistance changes upon exposure to $NO_2$, while semiconducting SWNTs are capable of detecting small concentrations of $NO_2$[36]. For the 1L and 6L sensors, the response almost linearly increases with the $NO_2$ concentration, as more electrons transfer from $Re_{0.5}Nb_{0.5}S_2$ to $NO_2$ when the $NO_2$ concentration is increased (Fig. 2a-b).

To identify the origin of the carrier type dependence on the $Re_{0.5}Nb_{0.5}S_2$ thickness, we compute the alignments of the energy levels with respect to the vacuum level up to 4 layers for the three lowest-energy AA´-type stackings. The alignment of the valence band and the conduction band edges with respect to the vacuum level is presented in Fig. 2c. Additionally, we can assume that defect states exist inside the gap in the experimental set up. They can act both as donors and acceptors depending on the chemical potential[37], which is determined by the work function of the metal contacts. In a simplified model, we can assume that all the defect states below the Fermi level of the metal contacts are occupied, and all the defect states above that level are unoccupied. Since our metal contacts have a work function of ~5 eV, these levels would promote n-type (p-type) behavior in the thinner (thicker) cases, because the change in the conduction band edge is smaller than that in the valence band edge as the thickness increases. This effect of the carrier type change with increasing thickness has been also observed in other 2D materials, such as $WSe_2$[18,19].

One of the most important characteristics of a chemical sensor is its selectivity to specific molecular species. Fig. 3a shows the responses of the 1L and 6L sensors to $NO_2$, $NH_3$, $CH_2O$, and $CO_2$ gases in dry air. Both sensors are strongly selective to $NO_2$ with only minimal responses to the other gases. For instance, responses of the monolayer sensor to $NH_3$ (1ppm), $CH_2O$ (2ppm), and $CO_2$ (5000ppm) are found to be ~1.25%, ~2.86%, and ~0.57%, respectively, while it shows a response of ~32.66% for $NO_2$ (1 ppm).

To understand the selectivity of the sensors toward $NO_2$, we computationally investigate the adsorption of various molecules on the monolayer crystal. We start with 16 random initial configurations for each molecule on $Re_{0.5}Nb_{0.5}S_2$ and allow them to relax to minimize the forces. We find that each molecule is physisorbed (Fig. 3b-e). The resulting adsorption energies are 0.22, 0.16, 0.19, 0.29 eV for $NH_3$, $CO_2$, $CH_2O$, and $NO_2$, respectively (Table S1), where the adsorption energy is defined as

$$E_{ad} = E_{substrate} + E_{molecule} - E_{substrate+molecule}. \qquad (1)$$

Since the adsorption energies of the molecules are within the same order of magnitude, the difference in the sensors' response to $NO_2$ compared to the other molecules is not due to potential differences in coverage. We then analyze the electronic structure of $Re_{0.5}Nb_{0.5}S_2$ with the adsorbed molecules. In Fig. 3f-i, we present the densities of states for the $Re_{0.5}Nb_{0.5}S_2$+adsorbed molecule systems. We observe that $NH_3$, $CO_2$ and $CH_2O$ contribute states that are deep in the valence and conduction bands of $Re_{0.5}Nb_{0.5}S_2$, whereas the $NO_2$ molecule contributes an unoccupied state 0.5eV above the valence band edge. If $Re_{0.5}Nb_{0.5}S_2$ were an ambipolar semiconductor experimentally, we would expect this state to act as an acceptor state and lead to p-type behavior. However, because of the observed n-type behavior of the monolayer device, we can assume that there are defect states in the gap that are filled up to a level closer to the conduction band edge. In this case, the $NO_2$ defect state in the gap would accept electrons from these states, reducing the n-type conduction, as experimentally observed. Contrarily, because the other molecules do not generate any gap states, they do not significantly modify the conduction of the system. The unoccupied $NO_2$ defect state remains close to the valence band edge for the thicker films that we computed (up to 4 layers). Therefore, we expect it to weaken conduction by electrons and strengthen conduction by holes for all thicknesses, as observed. The charge transfer from $Re_{0.5}Nb_{0.5}S_2$ to the molecule is also apparent in the projected densities of state (PDOS) plot in Fig. 3i, as the unoccupied in-gap state has non-zero projection onto the $Re_{0.5}Nb_{0.5}S_2$ states. This can also be observed in the real-space charge transfer plot (Fig. 3j-k). We note that $NO_2$ retains its paramagnetic character while adsorbed onto $Re_{0.5}Nb_{0.5}S_2$, and the current it contributes is expected to be spin-polarized. The calculated value of the electron transfer from $Re_{0.5}Nb_{0.5}S_2$ is equal to $0.10e$ for an $NO_2$ molecule, compared to $-0.01e$, $0.01e$, and $0.02e$ for $NH_3$, $CO_2$ and $CH_2O$, respectively. This is in line with previous studies that linked sensitivity and charge transfer in 2D materials[38,39].

All the sensing measurements so far described are performed in dry air. Since real-life sensors need to operate under typical atmospheric conditions, we test the monolayer sensor in the presence of humidity. Figure 4a exhibits the dynamic response of the monolayer sensor to $NO_2$ with concentrations ranging from 50ppb to 15ppm. Humidity not only does not deteriorate sensing properties of the monolayer device, but also is extremely beneficial for improving its recovery and response to $NO_2$ at room temperature.

For example, the recovery of the monolayer sensor after exposure to $NO_2$ (0.5ppm) is incomplete after being exposed to dry air for 360s (Fig. S6). However, a complete recovery is achieved when the sensor is exposed to air with 40% relative humidity (RH) in the same timeframe (Fig. 4a). The response and recovery times, defined as the time required to reach 90% of the resistance change upon exposure to and removal of $NO_2$ (15ppm), are approximately ~245 s and ~504 s, respectively, under 40% RH at room temperature. For comparison, an $NO_2$ sensor based on semiconducting SWNTs was demonstrated to detect 200ppm of $NO_2$ with a recovery time of 12 hours at room temperature and 1 hour at 200°C[36]. A metal oxide sensor based on $WO_3$ detected 500ppb of $NO_2$ with recovery times of 270s and 1350s at 300 and 150°C, respectively[12]. An $NO_2$ sensor based on monolayer $MoS_2$ was recovered to its initial state by leaving it in air for 12 hours at room temperature[23].

In addition to improving the recovery of the $Re_{0.5}Nb_{0.5}S_2$ sensor at room temperature, humidity also largely enhances its response to $NO_2$. For example, the monolayer sensor shows a response of ~195.53% for $NO_2$ (1ppm) in the presence of humidity (40% RH), compared to a response of ~32.66% in dry air. We also measure the response of the monolayer sensor to $NO_2$ under different humidity conditions (*e.g.* 20 and 80% RH) and find the 40% RH to be the optimum humidity condition while 20 and 80% RH still highly improve the recovery and response of the sensor at room temperature (Fig. S7). Testing the monolayer sensor under different humidity conditions, we see a decrease in the resistance of the device as the relative humidity increases from 0 to 80% (Fig. 4b). This is in contrast to the $MoS_2$ sensor where an increase in the resistance with humidity was observed[22]. Despite the fact that both the monolayer $Re_{0.5}Nb_{0.5}S_2$ and $MoS_2$ are n-type semiconductors, they interact with humidity very differently, analogous to the behavior of semiconducting metal oxide sensors. For instance, while a decrease in the resistance with humidity has been observed in n-type gas sensors such as $SnO_2$[40] and $ZnO$[41], $WO_3$, also n-type, has shown an opposite behavior[42].

In order to determine whether the monolayer sensor retains its selectivity to $NO_2$ in the presence of humidity, we test its response to other gas molecules in air with 40% RH. Fig. 4c displays the responses of the monolayer device to $NO_2$, $NH_3$, and $CH_2O$. Even in the presence of humidity, the monolayer sensor is still highly selective to $NO_2$ with insignificant responses to the other gases. The responses of the sensor to $NH_3$ (8ppm) and $CH_2O$ (20ppm) are ∼8.77% and ∼23.56%, respectively, compared to a response of ∼401.86% for $NO_2$ (5ppm).

To understand the origins of the improved response and recovery of the sensor in the presence of humidity, we first compute the optimal position of the $H_2O$ molecule to adsorb onto $Re_{0.5}Nb_{0.5}S_2$ (Fig. 4d-e). We then repeat the procedure to find the optimal configurations of higher $H_2O$ coverages (2, 3, and 4 molecules per 4×4 supercell). The adsorption energy per molecule for each coverage is given in Table 1. These calculations suggest that the energetic drive for the water molecules to adsorb onto $Re_{0.5}Nb_{0.5}S_2$ does not diminish even at very high coverage values. For each $H_2O$ coverage, we conduct a further search to find the lowest-energy configuration of the $NO_2$ adsorption (Fig. 4f-g). To elucidate the energetic drive for $NO_2$ on the $H_2O$-covered surface, we compute the adsorption energies using equation (1) where the substrate is defined as the combination of $Re_{0.5}Nb_{0.5}S_2$ and the already adsorbed $H_2O$ molecules (Table 1). We find that the $NO_2$ molecules lower their energy by adsorbing near the $H_2O$ molecules (by 0.05 eV for the lowest coverage), indicating a nonzero attraction between the adsorbed molecules. This attraction is due to the intrinsic and induced dipole moments of the molecules and the intermolecular charge transfer. The increase in the $NO_2$ adsorption energies with $H_2O$ coverage suggests that more $NO_2$ molecules adsorb onto $Re_{0.5}Nb_{0.5}S_2$ when there are more water molecules available.

The depletion of the $Re_{0.5}Nb_{0.5}S_2$'s charge also increases with $H_2O$ coverage, as listed in Table 1 ($\Delta\rho$ is defined as the total electron transfer to $Re_{0.5}Nb_{0.5}S_2$ from the adsorbed molecules). When a water molecule adsorbs onto $Re_{0.5}Nb_{0.5}S_2$, a small amount of charge transfer to the alloy occurs ($\Delta\rho = +0.01e$), as visualized in Fig. 4d-e. However, when the $NO_2$ molecule adsorbs at a nearby location on $Re_{0.5}Nb_{0.5}S_2$, the charge transfer between the alloy and the molecules is significantly modified (Fig. 4f-g), and $Re_{0.5}Nb_{0.5}S_2$ ends up losing more electrons ($\Delta\rho = -0.12e$) than it does due to the $NO_2$ molecule alone ($\Delta\rho = -0.10e$). The fact that the charge depletion of the alloy increases with higher $H_2O$ coverage can be visually observed by comparing the PDOS plots in Fig. 3i and Fig. 4h-i, as the projection of the unoccupied in-gap state onto the $Re_{0.5}Nb_{0.5}S_2$ states increases. The fact that both adsorption energy and charge transfer increase with humidity explains the improved response of the monolayer sensor to $NO_2$ in the presence of humidity.

Regarding the improved recovery rate, we propose a potential mechanism for the desorption of the $NO_2$ molecules in the presence of humidity driven by the intermolecular dipole-dipole interaction. First, we observe that both $NO_2$ and $H_2O$ are polar molecules due to their geometry, as opposed to the other molecules prominent in air, *i.e.* $N_2$ and $O_2$. If dipole–dipole interactions were the dominant cause of attraction between molecules, this qualitative difference between $H_2O$ and the other atmospheric gases would suggest that the $H_2O$ molecules attract the adsorbed $NO_2$ molecules while passing close to the alloy's surface, and "sweep" them away, whereas $N_2$ and $O_2$ do not. We find that although both $N_2$ and $O_2$ take on induced dipole moments, they do not cause an attraction comparable to that of H$_2$O (see Supporting Information and Table S2 for details). Therefore, we suggest that the significant improvement of the recovery rate with humidity may be due to the attraction between the H$_2$O and NO$_2$ molecules, which is significantly greater than those for N$_2$ and O$_2$.

In summary, we have demonstrated an NO$_2$ sensor based on a monolayer semiconducting alloy that has the advantage of room-temperature finite-humidity operation and ppb sensitivity. The sensor is highly selective to NO$_2$ with only minimal responses to other gases. In the presence of humidity, the sensor is highly reversible at room temperature with fast recovery time. The atomically-thin Re$_{0.5}$Nb$_{0.5}$S$_2$ sensor is flexible and optically transparent, making it attractive for a wide range of low-power sensor applications, such as in wearable electronics.

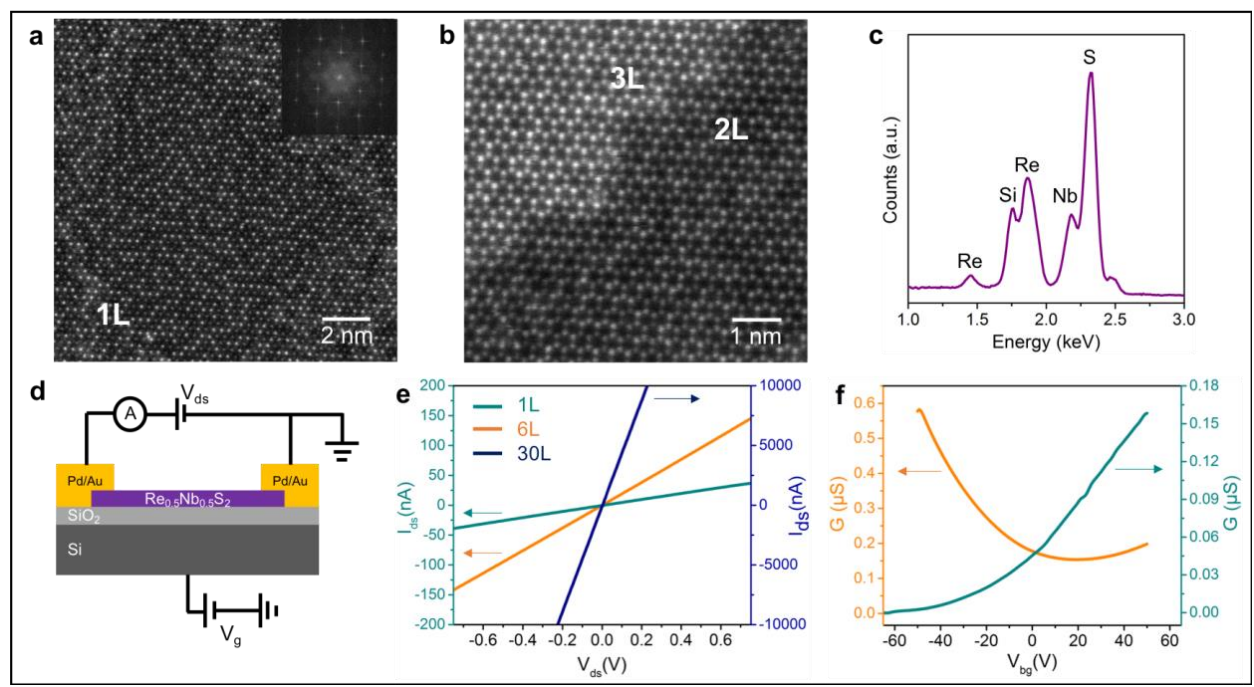

**Fig. 1 | Atomic structure, chemistry, and electrical characteristics of Re$_{0.5}$Nb$_{0.5}$S$_2$. a**, An ADF-STEM image of the monolayer Re$_{0.5}$Nb$_{0.5}$S$_2$ with the corresponding FFT (inset). **b**, An atomic-resolution ADF-STEM image of the bilayer and trilayer regions of Re$_{0.5}$Nb$_{0.5}$S$_2$ revealing its stacking order. **c**, EDS spectrum from a few-layer Re$_{0.5}$Nb$_{0.5}$S$_2$ crystal (see Fig. S1) showing peaks of Re, Nb, S, and Si (from the silicon nitride TEM grid). **d**, Schematic of the NO$_2$ sensors based on an FET device with a two-terminal back-gate configuration. **e**, Drain current ($I_{ds}$) as a

function of source-drain bias ($V_{ds}$) for the 1L, 6L, and 30L devices. **f**, Conductance (G) of the 1L, and 6L devices as a function of the gate voltage ($V_{bg}$).

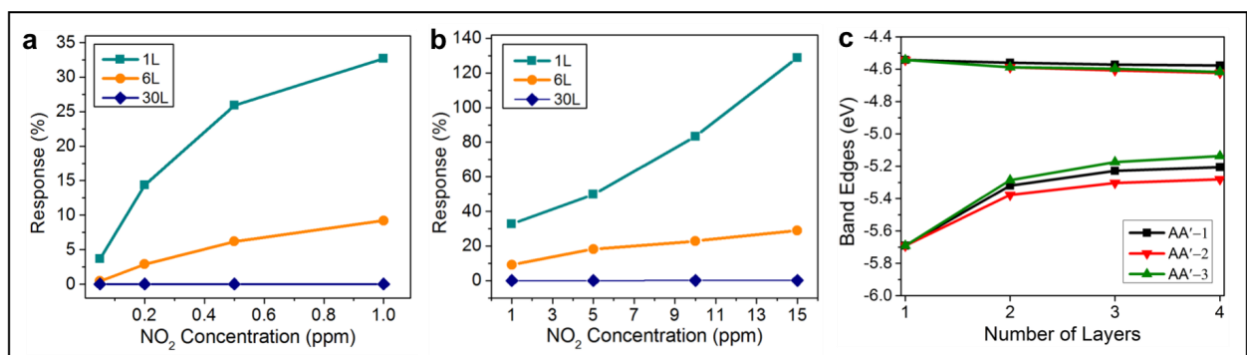

**Fig. 2 | Layer-dependent behavior of Re$_{0.5}$Nb$_{0.5}$S$_2$ NO$_2$ sensors**. The responses of the 1L, 6L, and 30L devices to different NO$_2$ concentrations raging from **a**, 50 ppb to 1 ppm and **b**, 1 to 15 ppm in dry air. Response is defined as $S = (R_g - R_{air})/R_{air}$, with $R_g$ and $R_{air}$ being the resistance of the device in target gas and air, respectively. **c**, The alignment of the valence band and the conduction band edges with respect to the vacuum level for three AA´-type stackings. Among several possible low-energy stackings of highly ordered model, we focus on the three lowest-energy AA´-type stackings (Fig. S4). See the energy level alignments with the adsorbed NO$_2$ molecule in Fig. S5.

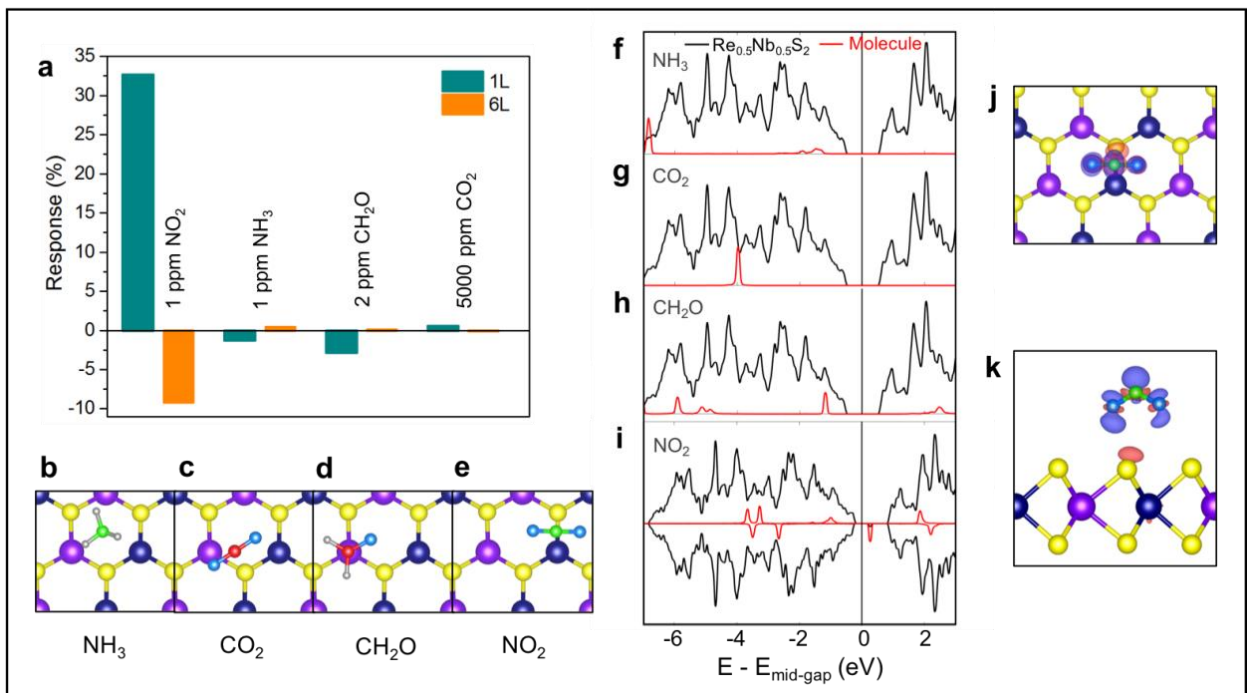

**Fig. 3 | Selectivity of the Re$_{0.5}$Nb$_{0.5}$S$_2$ sensors toward NO$_2$**. **a**, The responses of the 1L and 6L sensors to NO$_2$, NH$_3$, CH$_2$O, and CO$_2$ gases in dry air. **b-e**, NH$_3$, CO$_2$, CH$_2$O, and NO$_2$ molecules being physisorbed on the monolayer crystal. **f-i**, The densities of states for the Re$_{0.5}$Nb$_{0.5}$S$_2$ and the adsorbed NH$_3$, CO$_2$, CH$_2$O, and NO$_2$ molecule systems. Projections onto the atomic orbitals are

used to distinguish between the molecular states and the states in the substrate. The $NO_2$ molecule causes spin polarization in the system, and hence the two spins are plotted separately. **j-k**, The real-space charge transfer plot showing the charge transfer from $Re_{0.5}Nb_{0.5}S_2$ to the $NO_2$ molecule. Re: navy, Nb: violet, S: yellow, N: green, H: grey, O: blue, C: red.

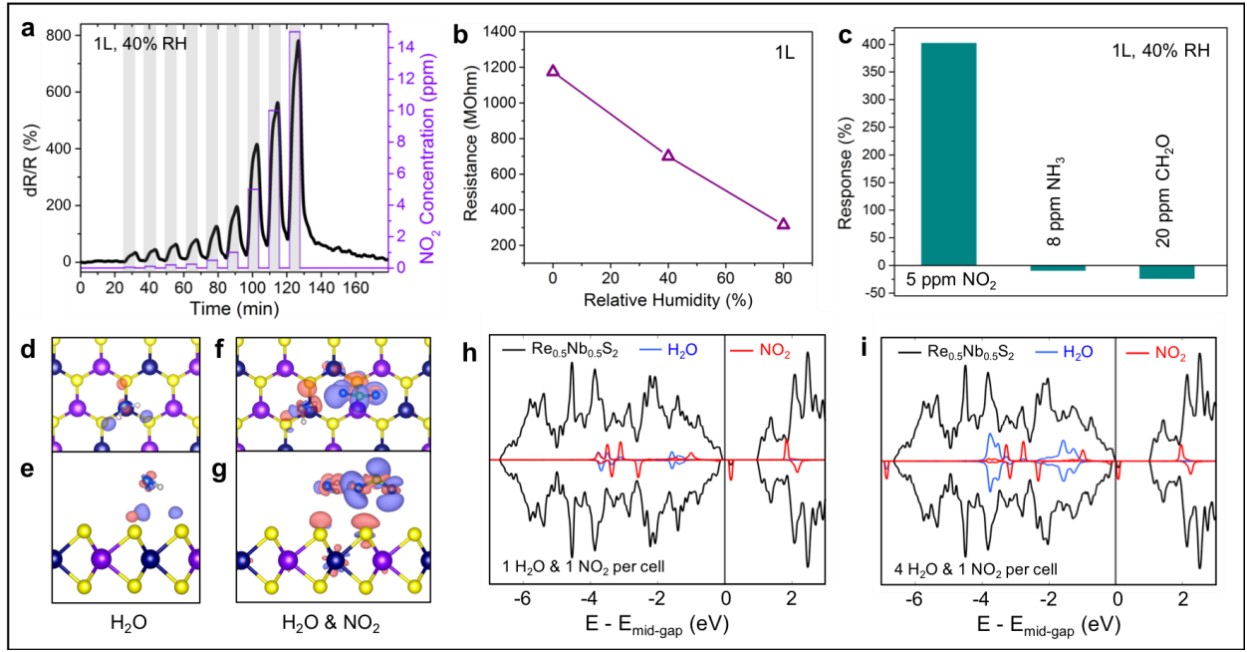

**Fig. 4 | Humidity impacts on the monolayer sensor. a**, Dynamic response of the monolayer $Re_{0.5}Nb_{0.5}S_2$ sensor to $NO_2$ with concentrations ranging from 50 ppb to 15 ppm under 40% RH condition at room temperature. **b**, Change in the resistance of the monolayer sensor as a function of the relative humidity. **c**, Responses of the monolayer device to $NO_2$, $NH_3$, and $CH_2O$ gases in presence of humidity (40% RH). **d-e**, The optimal position of the $H_2O$ molecule to adsorb onto the monolayer $Re_{0.5}Nb_{0.5}S_2$. **f-g**, The lowest-energy configuration of the $NO_2$ adsorption on the alloy for each $H_2O$ coverage. **h-i**, The PDOS plots for the $Re_{0.5}Nb_{0.5}S_2$ and the adsorbed $NO_2$ and $H_2O$ molecules, showing an increase in the charge depletion of $Re_{0.5}Nb_{0.5}S_2$ with higher $H_2O$ coverage. Re: navy, Nb: violet, S: yellow, N: green, H: grey, O: blue.

| $H_2O$ coverage per 4×4 cell | 0 | 1 | 2 | 3 | 4 |
|---|---|---|---|---|---|
| $E_{ad}(H_2O)$ (eV) |  | 0.16 | 0.16 | 0.15 | 0.14 |
| $E_{ad}(NO_2)$ (eV) | 0.29 | 0.34 | 0.41 | 0.42 | 0.51 |
| $\Delta\rho$ | -0.10e | -0.12e | -0.14e | -0.14e | -0.16e |

**Table 1 | Computed adsorption energies of $H_2O$ and $NO_2$ molecules on the $Re_{0.5}Nb_{0.5}S_2$ alloy, under various $H_2O$ coverages, and the charge transfer from the alloy to the adsorbed molecules.** The first (top) row lists the number of adsorbed $H_2O$ molecules in the 4×4 cell. The second row lists the adsorption energy of the $n_{th}$ $H_2O$ molecule in the cell. The third row lists the adsorption energy of the $NO_2$ molecule with $n$ $H_2O$ molecules already adsorbed onto $Re_{0.5}Nb_{0.5}S_2$. The entries in the fourth row correspond to the configurations in the third row, and list the electron

transfer to the alloy from the adsorbed molecules, after the adsorption of all the molecules ($NO_2 + n \times H_2O$). Negative $\Delta\rho$ values indicate that the alloy has lost electrons.

**Supporting Information**

The Supporting Information is available free of charge at http://pubs.acs.org. Methods, discussion on alignments of the energy levels with respect to the vacuum level (with the $NO_2$ molecule adsorbed), adsorption energies of different gas molecules on $NbS_2$, $Re_{0.5}Nb_{0.5}S_2$, and $ReS_2$, and improved recovery rate of the sensor in the presence of humidity, as well as additional details on chemical analysis, optical images of the samples, AFM measurement, and dynamic responses of the monolayer sensor to $NO_2$ in dry air and under different humidity conditions. (PDF)

**Acknowledgements**

This work was primarily supported by the U.S. Department of Energy, Office of Science, Office of Basic Energy Sciences, Materials Sciences and Engineering Division under Contract No. DE-AC02-05-CH11231, within the sp2-bonded Materials Program (KC2207), which provided for materials synthesis, chemical sensitivity tests, and atomic structure calculations. Additional support was provided by the National Science Foundation under Grant No. DMR-1807233 which provided for STEM measurements, and under Grant No. DMR-1926004 which provided for calculations of precise electronic structures. Computational resources were provided by the DOE at Lawrence Berkeley National Laboratory's NERSC facility and the NSF through XSEDE resources at NICS. We thank Sehoon Oh for fruitful scientific discussions.